# Supporting Orthographic Recall of Chinese Characters through Self-Generated Compositional Mnemonic Stories: A Mixed-Methods Pilot Study

# 借助自主生成的部件记忆故事促进汉字书写回忆：一项混合方法试点研究

Aaron F. Kurz
ETH Zurich (瑞士苏黎世联邦理工学院)
aakurz@student.ethz.ch, aaronfkurz@posteo.de

**Abstract:** Mnemonic studies on Chinese characters have mostly tested provided mnemonics against writing outcomes and self-generated mnemonics against meaning recall; whether writing one's own component-based story helps learners write the character has received little attention. I present Nuome, a tool that shows a character's component decomposition next to an editor in which learners write their own story, with live checks on whether the story covers every component and the character's meaning. In a within-subjects pilot, eight L2 learners each learned four self-chosen characters with the story builder and four with a read-only panel showing the same decomposition, with identical writing practice. Descriptively, partial-credit written-recall scores favoured the story builder by 18.8 percentage points immediately (N = 8) and 17.9 points after two days (N = 7); mean study time per character was 1.69 times as long (n = 6); usability was high (SUS 88.1). Interviews yielded three themes: participants felt that composing a story made them attend to components; knowing the components is not the same as being able to write them; and characters differ in how well they suit a story.

**摘要：**已有的汉字记忆法研究多检验研究者提供的记忆法对书写的效果，或学习者自主生成的记忆法对字义回忆的效果；学习者自己编写部件故事能否帮助他们写出汉字，较少受到关注。本文介绍工具 Nuome：它展示汉字的部件拆分，学习者在旁边的编辑器中用部件编写自己的记忆故事，工具实时提示故事是否涵盖了所有部件和整体字义。在一项被试内试点研究中，八名汉语二语学习者各自选择八个汉字，四个用故事编辑器学习，四个用只显示相同部件拆分的只读面板学习，两种条件的书写练习相同。描述性结果显示，故事编辑器条件下的部分计分书写回忆得分在即时测试中（N = 8）高出 18.8 个百分点，两天后（N = 7）高出 17.9 个百分点；每个汉字的平均学习时间

为对照条件的 1.69 倍（n = 6）；系统可用性量表（SUS）平均分为 88.1。访谈得出三个主题：学习者认为编写故事使他们更仔细地关注部件；知道部件并不等于会写部件；不同汉字是否适合编写故事因字而异。

## 1. Introduction

Writing Chinese characters is one of the hardest parts of Chinese for second-language (L2) learners: there are thousands of forms to memorise, and a character's form only has a weak, unreliable connection to its pronunciation. The usual method is rote copying. Handwriting practice does strengthen orthographic representations (Guan et al., 2011), but rote copying uses much practice time for limited gain. In a classroom crossover study, hand-copying took about a third of practice time, lowered word recognition compared with other uses of that time, and its handwriting advantage was no longer detected at a one-week posttest (Lu et al., 2019).

Most characters are built from a limited set of recurring components. In one study, explaining radicals helped L2 learners learn new characters (Taft & Chung, 1999), and learners rate component-based strategies as useful (Shen, 2005). Components can also be the material for a mnemonic. The generation effect, the finding that information people generate themselves is remembered better than information they only read (Bertsch et al., 2007; Slamecka & Graf, 1978), suggests that learners who build their own mnemonic from a character's components may remember the character better than learners who only study the components. Evidence on whether this holds for writing is limited. This paper addresses the question with Nuome, a tool in which learners write their own mnemonic story from a character's components, and a mixed-methods pilot study with eight L2 learners. The research questions were:

**RQ1.** Does building a component-based mnemonic story lead to higher written recall at an immediate and a two-day delayed test than studying the same decomposition without a story, when both conditions receive identical writing practice?

**RQ2.** How usable is the story builder, and what do learners find helpful or frustrating?

## 2. Related Work

Wang and Thomas (1992) compared imagery mnemonics supplied by the experimenter with rote copying for recall of character meanings; the advantage did not last beyond the immediate test. They suggested that self-generated mnemonics might last longer, but could not test this. Kuo and Hooper (2004) compared five ways of coding characters with 92 high-school learners. Learners who generated their own mnemonics outperformed the visual, verbal, and translation groups, but they also spent more time on task, and the test measured meaning, not writing.

The studies below that measured writing used provided mnemonics. Chang et al. (2019) found that key-image mnemonics designed by the researchers led to better form reproduction than etymological explanation or stroke-sequence copying. Chang et al. (2022) found that key images combined with a short verbal formula gave the highest immediate writing accuracy for visually similar characters, but no condition differed at a one-week posttest. Leminen and Bai (2023) found that adding visual mnemonics to decomposition and writing practice improved retention among 30 near-beginners. Xu and Padilla (2013) found that meaningful interpretation and chunking (MIC), which includes mnemonic techniques and interpreting components, improved immediate learning and one-week retention compared with rote repetition; MIC does not require learners to write their own story, and its outcomes combine several aspects of character knowledge. *Remembering the Hanzi* (Heisig & Richardson, 2009) has learners invent stories that link a character's parts to its meaning in order to write it, but I found no controlled evaluation of it with a scored writing outcome.

Table 1 arranges these studies by source of the mnemonic and outcome. It is not an exhaustive review, and none of the studies in it tested self-generated mnemonics against written recall. This study tests whether writing one's own mnemonic on top of studying the decomposition helps, not whether self-generated mnemonics beat provided ones.

Table 1. Mnemonic studies on Chinese characters discussed in this paper, by source of mnemonic and outcome

| **Source of mnemonic** | **Meaning recall and recognition** | **Written recall** |
|---|---|---|
| Provided | Wang & Thomas (1992); Kuo & Hooper (2004) | Chang et al. (2019, 2022); Leminen & Bai (2023) |
| Self-generated | Kuo & Hooper (2004) | This study |

## 3. The Mnemonic Builder

Nuome (Kurz, 2026) is a web application that uses a dictionary of over 9,500 characters with pinyin, English definitions, and component decompositions from the open-source Make Me a Hanzi dataset (Kishore, 2016). The decompositions describe the modern printed form, not etymology. The story builder (Figure 1) shows the character, its meanings, and a tree of its components with pinyin and glosses next to a free text editor. Tapping a component or meaning inserts it into the story as a labelled chip. A coverage check lists components not yet used in the story (a component counts as covered if it, its parent, or all of its sub-components are used), and a meaning check tests whether any English gloss of the character appears in the story.

After the story, the learner uses a five-stage writing trainer: copying each first-level component (the character's immediate parts) from an animated stroke-order model into a separate grid, writing the components from memory, assembling them into the character's layout, writing the whole character once with guidance, and writing it twice from memory, on a 米字格 grid with a finger (on a touchscreen, as in this study) or a mouse.

## 4. Method

**Participants.** Eight L2 learners of Chinese (four women, three men, and one participant of another gender; aged 18 to 35) were recruited through a university language centre. All had taken formal classes, for between about six months and more than two years. Advanced learners and complete beginners were excluded, since participants needed eight characters they could not write and some component knowledge. Six rated their reading as beginner level; two could not yet write characters by hand. All gave written informed consent. One did not complete the delayed test ($N = 7$ for delayed analyses). The study was approved by the ETH Zurich Ethics Commission (project 26 ETHICS-309).

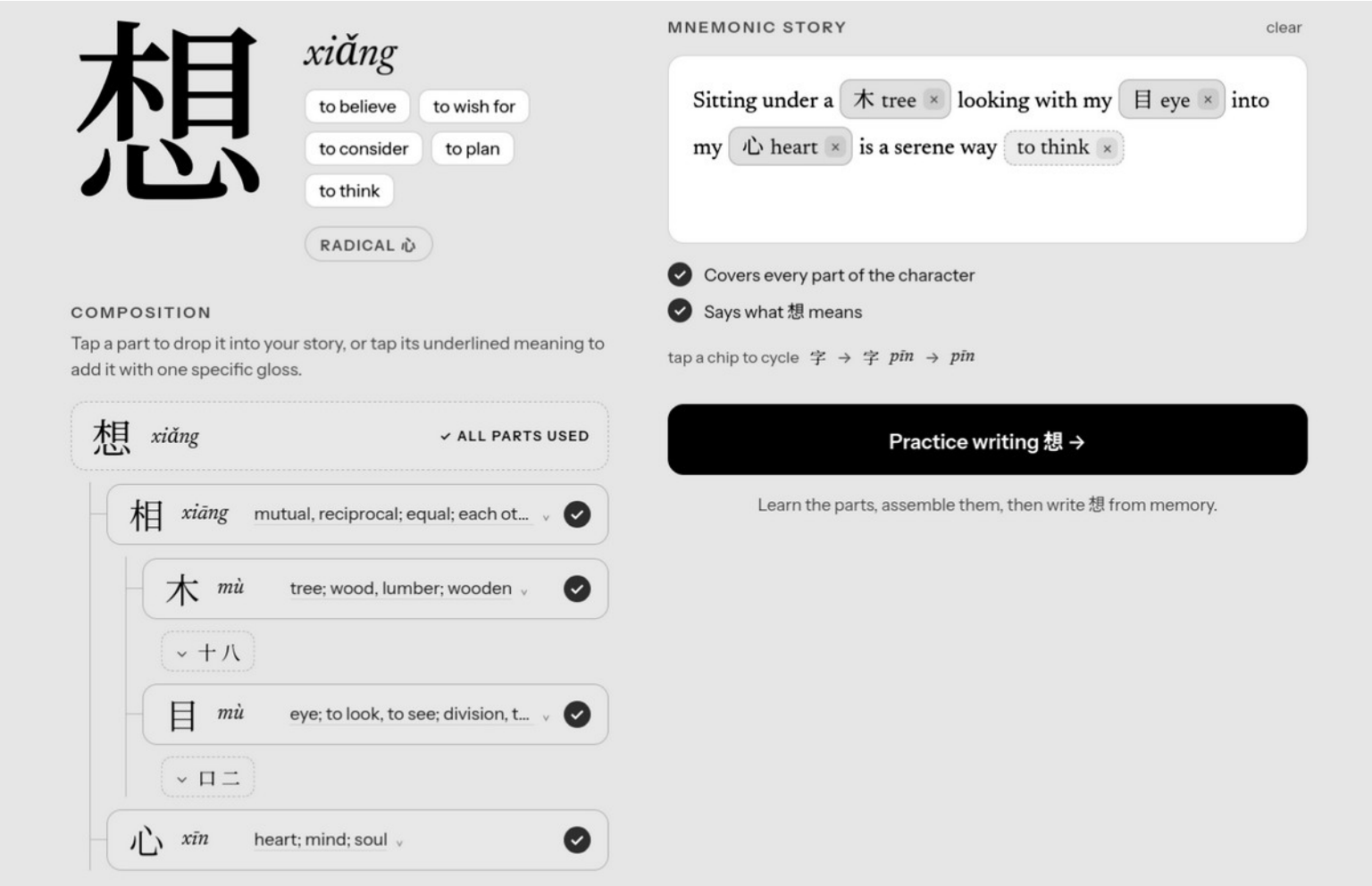

Figure 1. The story builder with 想 *xiǎng* ("to think"): component tree (left) and an example story with inserted components and the two checks (right).

**Design and procedure.** Condition was assigned per character within participants. Each participant chose eight characters that they could not yet write, each with at least two components. The tool ranked the chosen characters by stroke count, paired neighbours, and randomly assigned one of each pair to each condition, then interleaved the two sets, with a random choice of which condition came first (the alternation held throughout the session); participants were not told the assignment. In the mnemonic condition, participants wrote a story with the story builder; in the control condition, a read-only panel showed the same components, pinyin, and glosses. Both conditions then used the same trainer. Study time was not equalised, because the pilot asked whether writing a mnemonic helps at all, and filling the control time with extra writing would have added handwriting practice (Guan et al., 2011). The study times reported below were measured from screen recordings.

After a distractor task, participants took a written recall test: they saw the pinyin and English gloss of each character in random order and wrote the character by hand. They then completed the System Usability Scale (SUS; Brooke, 1996), the interest/enjoyment subscale of the Intrinsic Motivation Inventory (IMI; McAuley et al., 1989), ratings of ease of learning per condition, four 5-point items on specific features (reported in Section 5), and a question on whether they had made up stories for control characters, followed by a semi-structured interview. Two days later, they took the same test remotely and unsupervised.

**Scoring and analysis.** The outcome is a partial-credit score: each character was scored 1 if correct, 0.5 if at least one first-level component was written correctly, and 0 if wrong or blank. Scores are reported as percentages of the maximum, so 50% does not mean that half of the characters were fully correct. The author scored all responses while the scoring interface hid each character's condition. I report how many participants scored higher in each condition, a Wilcoxon signed-rank test with exact *p* values read descriptively, and the rank-biserial correlation (Kerby, 2014). The interviews were audio-recorded and transcribed, and the author coded the transcripts alone, inductively and mostly at a semantic level, following reflexive thematic analysis (Braun & Clarke, 2006). While coding, the author looked in particular for material on the questions the study was built around: whether stories helped or felt forced, strategy carryover, the component breakdown, usability, and what participants recalled at test. Quotations were translated from German by the author.

## 5. Results

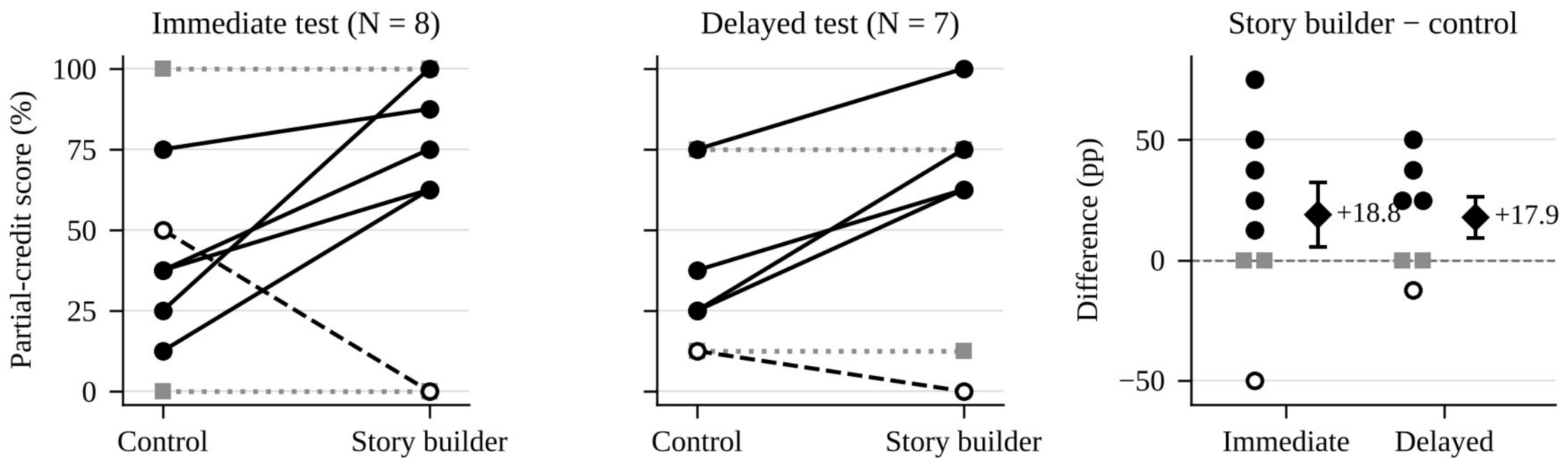

Figure 2. Scores per participant (left, centre) and story builder minus control differences (right; diamonds: mean ± 1 SE). Solid or filled: higher with story builder; dashed or open: higher with control; grey: tie.

**Written recall and study time.** Figure 2 shows each participant's scores. Mean scores with the story builder and with control were 60.9% (*SD* = 40.3) and 42.2% (32.7) at the immediate test, a difference of 18.8 percentage points, and 55.4% (36.0) and 37.5% (27.0) at the delayed test, a difference of 17.9 points. At the immediate test, five of eight participants scored higher with the story builder, one scored lower, and two tied ($W = 4.5$, exact two-sided $p = .25$; $r = .57$). At the delayed test, four of seven scored higher, the same participant as before scored lower, and two tied ($W = 1.0$, $p = .125$; $r = .87$). The delayed result rests on seven participants. Mean study time per character, from the start of one character to the start of the next, was 216.3 s (*SD* = 92.7) with the story builder and 128.3 s (43.8) with control, a ratio of 1.69; SDs are across the 24 characters per condition from six participants. Timing data are available for six of the eight participants because of a file corruption issue, and all six took longer with the story builder.

**Questionnaires.** The mean SUS score was 88.1 (*SD* = 5.9), above the average of about 68 reported across many studies (Lewis, 2018), and mean IMI interest/enjoyment was 5.77 of 7. Story-builder characters were rated somewhat easier to learn than control characters (4.38 vs. 3.75 of 7). On four statements rated from 1 (strongly disagree) to 5 (strongly agree), mean agreement was 4.38 for “The component breakdown helped me understand the character,” 3.63 for “The story/mnemonic step helped me remember how to write the character,” 3.13 for “I could recall the character by recalling its story,” and 2.50 for “The mnemonic felt natural rather than forced.” Two participants made up stories for control characters. Usability problems raised in the interviews were minor (scrolling on phones, inserted words not fitting the learner's grammar, tap targets too close together), and participants asked for stroke checking and a way to mark components as already known.

**Theme 1: Participants felt that composing the story made them attend to components.** Participants located the benefit in looking closely at components while building the story, not in recalling the story. P5: “you really have to focus on the radicals. […] When writing it down, I imagined the story less—but I did think: right, that was that radical and that was that radical.” P1, in contrast, found the step unnecessary while doing it, but at test noticed that recalling the story itself helped (“ah right, that was the guy playing the flute energetically”).

**Theme 2: Knowing what goes where is not knowing how to write it.** Some participants knew the components from the story but could not write them: “I know it's bamboo, but I don't know how to write it” (P2). P4, who knew many components, did not report this problem.

**Theme 3: Some characters suit a story and others do not.** P1 and P6 each said that stories work less well for characters with a phonetic component. P4 said stories work for characters that contain a scene, such as a “broken heart.”

## 6. Discussion

Partial-credit written-recall scores favoured the story builder at both tests. With eight participants and no statistically significant difference, this is directional evidence only. The direction matches Kuo and Hooper's (2004) self-generation advantage, although they measured meaning recall, not writing, and a two-day delay cannot show whether the difference lasts longer.

What produced the difference is not clear. Participants attributed it to the attention to components that writing a story required, which fits a levels-of-processing account (Craik & Lockhart, 1972) and the idea of desirable difficulties, in which a step that feels costly during learning pays off later (Bjork, 1994). Because the mnemonic condition took 1.69 times as long, the design cannot separate the effects of generating the story oneself, of having a narrative at all, of the extra attention, and of the extra time. The interviews also suggest limits (Themes 2 and 3), although component familiarity was not measured.

For teaching, these findings tentatively suggest using learner-built stories selectively, for characters whose components the learner can already write and whose meaning lends itself to a scene. Learners may need practice writing common components first, and to be told why the step is worth the time, since it can feel redundant while they do it. The story builder could also be a targeted step in a spaced-repetition flow, used only for characters a learner keeps forgetting ("leeches" in tools such as Anki). This proposal is untested.

**Limitations and future work.** The sample is small, study time was not matched, two participants carried the story strategy over to control characters, scoring used a single rater, the delayed test was unsupervised and short, and the delayed and timing results rest on seven and six participants. A follow-up should compare self-written with provided stories and with a control that requires similar attention to components without a narrative, use a between-subjects design and a longer delay, and measure each learner's familiarity with the components.